\documentclass[aps,showpacs,showkeys,twocolumn,groupedaddress]{revtex4}
\usepackage[T1]{fontenc}
\usepackage{url}
\usepackage{graphicx}

\begin{document}
\title{Wringing out DNA}
\author{Timoth\'ee Lionnet$^1$}
\email[]{lionnet@lps.ens.fr}
\author{Sylvain Joubaud$^2$}

\author{Richard Lavery$^3$}
\author{David Bensimon$^1$}
\author{Vincent Croquette$^1$}
\affiliation{$^1$ Laboratoire de Physique Statistique, CNRS UMR 8550, Ecole Normale Sup\'erieure, 24 rue Lhomond 75005 Paris, France\\
$^2$ Laboratoire de Physique
CNRS UMR 5672, Ecole Normale Sup\'erieure de Lyon,
46 allée d'Italie 69364 Lyon Cedex 07, France \\
$^3$ Laboratoire de Biochimie Theorique, CNRS UPR 9080, Institut
de Biologie Physico-Chimique, 13 rue Pierre et Marie Curie,
Paris 75005, France}
\date{\today}

\begin{abstract}
The chiral nature of DNA plays a crucial role in cellular processes. Here we use magnetic tweezers to explore one of the signatures of this chirality, the coupling between stretch and twist deformations. We show that the extention of a stretched DNA molecule increases linearly by 0.42 nm per excess turn applied to the double helix. This result contradicts the intuition that DNA should lengthen as it is unwound and get shorter with overwinding. We then present numerical results of energy minimizations of torsionally restrained DNA that display a behaviour similar to the experimental data and shed light on the molecular details of this surprising effect.
\end{abstract}

\maketitle

The helical structure of double-stranded DNA (dsDNA) results in very specific mechanical properties. These affect the function of nucleic acids, as they determine the accessibility of the genetic material to the proteins that process it. Micromanipulation techniques developed over the past decade have made it possible to manipulate a single DNA molecule, and thus to extensively study its mechanical response \cite{strickReview2000,allemandCurrOpinStructBiol2003,laveryJoPCondMat2002}. 
In the so-called entropic regime ($F\ \lesssim$  10 pN), stretched dsDNA behaves as a uniform, semi-flexible rod whose extension results from a balance between the entropy of bending fluctuations and the work performed by the stretching force \cite{smithScience1992,bustaScience1994,markoMacromolecules1995,bouchiatBiophysJ1999}. 
In the very low force regime (typically $F <$ 0.4 pN), twisted DNA behaves as an isotropic rod. Its torsional energy increases up to a threshold where the molecule buckles to form superhelical loops called plectonemes \cite{strickScience1996,strickBiophysJ1998,vologoBiophysJ1997,morozPNAS1997,bouchiatPRL1998,zhouPRE2000}. At higher forces (and consequently higher torques), DNA chirality affects its response to twist. The molecule undergoes structural transitions away from its native conformation in solution, {\em B}-DNA. When the torque is negative, {\em i.e.} opposite to the handedness of the double helix, DNA denatures: its unwound strands locally separate. When the torque is positive, the molecule adopts a highly overwound state, {\em P}-DNA \cite{allemandPnas1998, strickBiophysJ1998, bryantNature2003}.

As a result of its helical structure, a DNA molecule is expected to display a coupling between its extension $L$ and its degree of twist $Tw$. In other words, one expects the extension of DNA in its native structure to be an asymmetric function of twist. The response of DNA to both twist and stretch has been studied in the framework  of continuum elasticity \cite{markoEurophysLett1997,kamienEurophysLett1997,ohernEurPhysJB1998}. In that approach, the energy per unit length $E/L_0$ is a quadratic function of its relative change in extension $\epsilon = L/L_0 - 1$ and twist: $\sigma = Tw/Tw_0-1 = n/Tw_0$, (where $L_0$ and $Tw_0$ are the contour length and natural twist of the double helix and $n$ is the number of turns added to a twisted DNA):
$$
E/l_0 = \frac{4\pi^2k_BT}{p^2}(C \sigma^2 + B \epsilon^2 + 2 D \epsilon\sigma) - F\epsilon
$$
where $p = L_0/Tw$ = 3.6 nm is the DNA pitch and $C \approx$ 100 nm, $B\approx$ 78 nm and $D$ are respectively the DNA torsional, stretch and twist/stretch moduli \cite{charvinCP2004,strickReview2000}. By minimizing $E$ with respect to $\epsilon$ at fixed $\sigma$, one finds that the extension should vary linearly with the twist: $\epsilon - \epsilon_{\sigma = 0}= -D\sigma/B$, equivalent to $\Delta L = -pDn/B$. Values of $D$ in the 12-20 nm range have been previously extracted from single molecule results \cite{markoEurophysLett1997,kamienEurophysLett1997}. This corresponds to a decrease in length of the DNA molecule of 0.6-0.9 nm per added turn. However, these estimates must be taken with care as they were extracted from data pertaining to DNA adopting different structures, not necessarily the native {\em B}-DNA form.

\begin{figure}
\includegraphics[width=7cm]{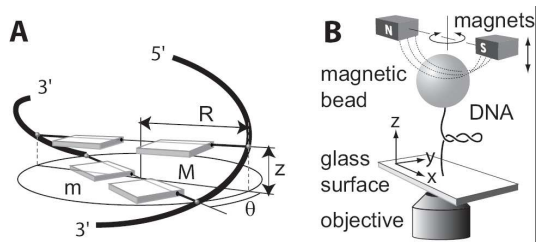}
\caption{\label{setup} A: Two adjacent base pairs are schematically drawn as two pairs of plates, connected by the two backbones (thick black lines). Notice the negative value of the shift in {\em B}- DNA, {\em i.e.} the displacement of the base pair from the helical axis towards the minor groove (marked by 'm'). 'M' marks the major groove. B: Principle of the magnetic tweezers (not to scale).}
\end{figure}

In the absence of a reliable twist-stretch estimate, elasticity theory needs to be backed by a complementary, atomic-scale approach, in order to predict the value of the phenomenological parameter $D$. Geometry alone provides a first "naive" estimate, if we assume that the double helix radius $R$ = 1 nm and the arc length of the backbones $N_{bp} \sqrt{R^2\theta^2 + z^2}$ are fixed ($N_{bp}$ being the total number of base pairs, $z = L/N_{bp}$ the rise per base pair -the difference between adjacent base pairs along the helical axis- and $\theta = 2\pi Tw/N_{bp}$  the twist per base pair -the angle formed by adjacent base pairs around the helix axis- see Fig.\ref{setup}A). The length of the molecule should then increase as it is unwound: $dL/dn = -4\pi^2 R^2/p \; \approx \;$ - 11 nm/turn (or $D$ = 240 nm). A finer approach based on Monte Carlo simulations suggests a similar estimate \cite{mergellPRE2003}. Elastic constants calculated from Molecular Dynamics (MD) simulations yield $D$ values ranging from 4 to 18 mn for different base sequences, which implies an average value of dL/dn = -0.5 nm/turn \cite{lankasJMB2000}. All these estimates, in spite of important differences in their magnitude, agree on the twist-stretch coupling sign: DNA should lengthen as it is unwound, as suggested by the simple picture of wringing out a floorcloth.

We now describe magnetic tweezer experiments on single DNA molecules that allow measurement of the value of the twist-stretch coupling of {\em B}-DNA. While its sign is opposite to the "naive" expectations previously mentioned, it is however in good agreement with atomic-scale numerical modeling of the response of DNA to twist. The modeling results further suggest that response of {\em B}-DNA to torsion illustrates the same mechanical coupling as seen in the transition between {\em B}-DNA and the form of DNA observed under low hydration conditions, {\em A}-DNA \cite{saenger}.

{\it Magnetic tweezer experiments -- } Magnetic tweezers allow to monitor the change in extension of a single dsDNA molecule as it is twisted \cite{strickScience1996,gosseBiophysJ2002}. In a custom-built flow chamber, we tether a single DNA molecule between a glass surface and a superparamagnetic bead (1.4 $\mu$m radius, Dynal). A pair of permanent magnets placed above the sample generate a constant, vertical stretching force on the bead and lock its rotational motion around the vertical axis. The exerted force $F$ and the rotation $n$ of the bead can be set by respectively translating and rotating the magnets (Fig.\ref{setup}B). The bead is imaged at video rate (60 Hz) through a 100x oil immersion objective, with a typical resolution of 2 nm. The force $F$ exerted on the bead is computed from measurements of $\langle \delta x^2\rangle$ and $L$ using the equipartition theorem: $F=k_BTL/\langle \delta x^2\rangle$ \cite{gosseBiophysJ2002}.

\begin{figure}
\includegraphics[width=7cm]{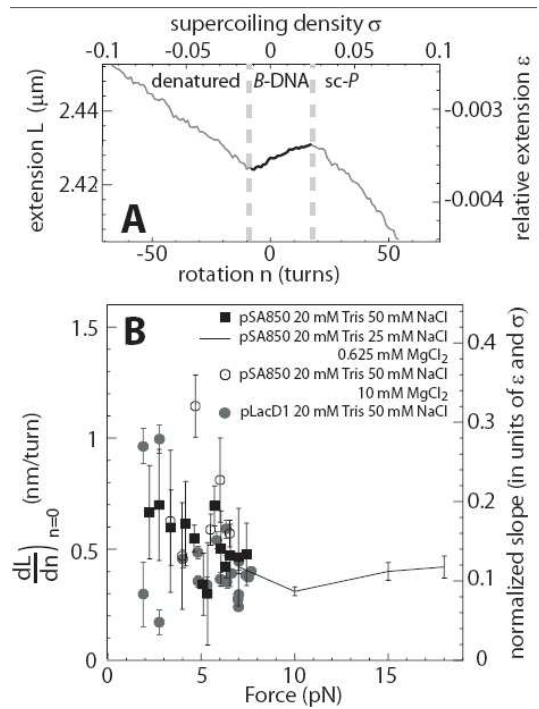}
\caption{\label{hatcurve_and_slope}A: Extension $L$ versus rotation $n$ curve of DNA (pLacD1, 7.4 kbp; 20 mM Tris pH 7.9, 25 mM NaCl, 0.625 mM MgCl$_2$, 0.1 $\%$ w/v BSA; $F$ = 7 pN). In the {\em B}-DNA stability region (black overlay), the curve displays a linear behaviour, with a slope of 0.28 nm/turn. B: Effect of the buffer conditions, exerted force and DNA sequence content on the twist-stretch coupling. Since the slope value does not display any significant variation, we extract its average value (S.D.) in nm/turn from each set of data: solid squares: 0.52 (0.13); black line: 0.39 (0.05); open circles: 0.70 (0.24); solid gray circles: 0.43 (0.21).}
\end{figure}

The curves $L(n)$ are obtained by rotating the end of a DNA molecule under a constant force $F$  and averaging the measured extension $L$ for each value of $n$ over typically 128 points. At high enough $F$ and for small $n$ values (typically -0.01 < $\sigma$ < 0.02), rotation translates into a change of the DNA twist. This approximation, which neglects relaxation of torsional constraint through bending fluctuations ("writhe"), is correct (when $ F \gtrsim$ 2 pN) within a 10 \% margin \cite{morozPNAS1997}. However, for larger $|\sigma|$ values, buckling and/or structural transitions invalidate this relationship. This is illustrated in Fig.\ref{hatcurve_and_slope}A, where dsDNA is stable in its {\em B}- form for small torsional strains (black overlay, -0.01 < $\sigma$ < 0.02). Storing twist energy through small deformations around the relaxed {\em B}- conformation becomes unfavorable under higher torsional strain, and dsDNA undergoes a transition to supercoiled {\em P}-DNA (sc-{\em P}) in the positive supercoiling region ($\sigma$ > 0.02), whereas strong negative supercoiling conditions induce local denaturation of the molecule ($\sigma$ < -0.01) \cite{strickBiophysJ1998,sarkarPRE2001}. The twist-stretch coupling measures the change in extension for small variations in twist. It is thus deduced from the slope of $L(n)$ at $n \sim $0: $\frac{dL}{dn})_{n=0}$ = 0.28 $\pm$ 0.04 nm/turn (Fig.\ref{hatcurve_and_slope}A) equivalent to a twist-stretch modulus $D$ = -6.1 $\pm$ 0.9 nm. Notice that as B-DNA is overwound its extension increases, in contradiction with one's naive intuition.

%\pagebreak

%\pagebreak

The twist-stretch coupling is neither force- nor [Mg$^{2+}$]- dependent (within the 2-20 pN and 0-10 mM range explored (Fig.\ref{hatcurve_and_slope}). We did not observe any sequence dependence, as two unrelated molecules with different sequence contents (pLacD1: 7.4 kbp, 43 $\%$ GC; pSA850: 3.6 kbp, 48 $\%$ GC) yield similar values. By averaging all the values obtained in the different conditions, we obtain a slope of 0.42 $\pm$ 0.2 nm/turn, and thus $D$ = -9.1 $\pm$ 4 nm.

{\it Molecular modeling calculations -- } In the absence of a satisfactory theoretical description, we decided to investigate the atomic details of dsDNA response to twist in order to understand our experimental results. This has been done by performing energy minimization of DNA under conditions mimicking our experiments.

The modeling calculations were carried out with JUMNA \cite{laveryCompPhysComm1995} using the AMBER parm 98 force field \cite{cheathamJBiomolStructDyn1999}. DNA was modeled as a helically symmetric polymer with a mononucleotide or dinucleotide symmetry repeat according to the nature of the base sequence. Solvent and counterion effects were modeled using a sigmoïdal distance dependent dielectric function and reduced phosphate charges (-0.5 $e$). Calculations made with a generalized Born solvent model \cite{tsuiJACS2000} gave very similar results. Sequence effects were investigated by making calculations for all dinucleotide repeating sequences, (AA)$_n$, (GG)$_n$, (AT)$_n$, (CG)$_n$, (AC)$_n$ and (AG)$_n$, and for an "average" sequence composed of equal contributions from each of the four standard base pairs, obtained using the multicopy approach ADAPT \cite{lafontaineBiophysJ2000}. For each sequence, the twist per base pair was fixed at values ranging from 32.5$^\circ$ to 40.5$^\circ$ (at intervals of 0.25$^\circ$). The DNA structure was then energy minimized and its helical conformation was analyzed. The effects of tension were studied by applying equal and opposite forces to both ends of each strand of the double helix \cite{laveryGenetica1999}.

\begin{figure}
\includegraphics[width=7cm]{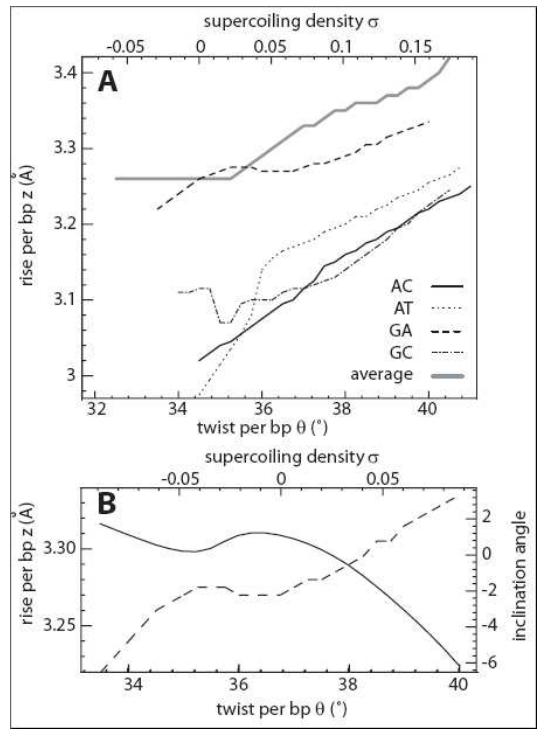}
\caption{\label{simu}A: Twist-stretch coupling extracted from energy minimizations ($F$ = 0), for different dinucleotide sequences. All curves display a linear behaviour around the relaxed state. Fitting each curve in its linear region to a line yields the value of the slope in nm/turn: 0.9 (average), 1.15 (GC), 0.97 (AT), 1.33 (AC) and 0.76 (GA). $\sigma$ = 0 refers to the relaxed twist per base pair for the average sequence (34.5$^\circ$). B: variation of the rise (dashed) and inclination (solid) as functions of the induced twist for the GA sequence. $\sigma$ = 0 refers to the relaxed twist per base pair for the GA sequence  (36.75$^\circ$).}
\end{figure}

In the absence of an exerted force, the base pair extension, or rise, $z$ is a linear function of the twist per base pair $\theta$ in the positive supercoiling region, {\it i.e.} when the twist is greater than its relaxed state value $\theta_0 \approx$ 34.5$^\circ$ (Fig.\ref{simu}A). All tested sequences have comparable slopes, typically %0.026 \AA/$^\circ$ = 0.9 nm/turn 
0.9 nm/turn (thus $D$ = -20 nm). In the negative supercoiling region, the curves display different behaviours that cancel out to produce a flat line in the case of the "average" sequence. Exerting a force on the molecule slightly decreases the slope of the rise versus twist curve. At 6 pN (18 pN), the rise increases by 0.68 nm/turn (0.58 nm/turn), equivalent to $D$ = -15 nm (-13 nm). 
 
%\pagebreak

Other basepair parameters, such as inclination, roll, etc \cite{dickersonNAR1989} were monitored as the twist is varied. As shown in Fig.\ref{simu}B, increasing rise is tightly coupled to more negative inclination ({\em i.e.} a counterclockwise rotation of the base pairs viewed from the minor groove of the double helix). The absence of significant rise variation observed for energy minimizations of a DNA molecule with locked base pair inclination further illustrates this tight coupling (data not shown). Other structural parameters display a significant change in this range of twist values: the shift becomes less negative with increasing twist, with a slope of 0.38 \AA /degree (shift, or, more formally, the $X$ displacement is the displacement of the base pairs from the helical axis perpendicular to their long axis, and is positive for displacements towards the major groove); the sugar phase angle \cite{saenger}) 
increases by 2.5 degree per degree of twist; finally, the double helix diameter decreases by -0.44 \AA  /degree.

{\it Concluding remarks -- }
The magnetic tweezer experiments presented in this work provide clear evidence of a linear twist-stretch coupling in {\em B}-DNA, in the regime of small, physiological torsional strains ($|\sigma| < 0.05$). 
The sign and magnitude of this effect differ from naive theoretical predictions. However, they are in good agreement with the present molecular modeling calculations of the response of {\em B}-DNA to limited changes in the twist.

A few details remain to be adressed. Calculations predict that $D$ should decrease with increasing $F$. Experimental data lack the necessary resolution at low force to observe such an effect. In addition, the range of supercoiling values in which this effect is observed differs between experiments and calculations. Energy minimizations include neither the possible structural transitions to different DNA states (denatured or P) nor sense of buckling transition. This explains why a larger $\sigma$ range can be studied than in the tweezer experiments. However, it is not clear why linear, positive twist-stretch coupling is only observed for positive supercoiling in the calculations, whereas experimentally, such a behaviour is observed for both positive and negative supercoiling.

The difference between our results and prior $D$ estimates based on single-molecule experiments is most likely due to the presence of extreme structural transitions in the data previously exploited \cite{markoEurophysLett1997, kamienEurophysLett1997}. This suggests that structural transitions exploit very different atomic mechanisms than fluctuations within the {\em B}-DNA stability region. Work remains to be done to clearly relate the present results to MD simulations \cite{lankasJMB2000,lankasBiophysJ2003}. 

\begin{figure}
\includegraphics[width=7cm]{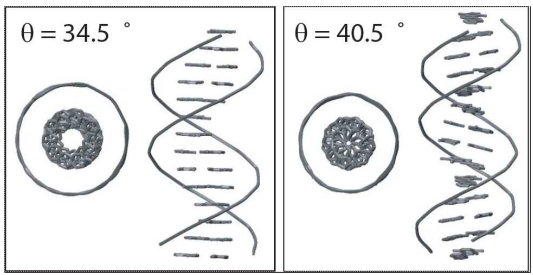}
\caption{\label{structure} Axial and lateral views of the structures obtained by energy minimizations (14 base pairs; "average" sequence). Only the phosphodiester backbones and base pairs are represented. Left: relaxed {\em B}-DNA state, $\theta$ = 34.5$^\circ$. Right: {\em B}-DNA under positive supercoiling ($\theta$ = 40.5$^\circ$) displays a smaller diameter, an increased length, a negative base pair inclination, and a less negative shift value.}
\end{figure}

The present numerical results yield evidence for the molecular mechanism of the observed effect: the decrease in rise as the molecule is unwound is coupled to an inclination of the basepairs towards positive values, an increase in the diameter, more negative shift values, and a decrease of the sugar phase angle (Fig.\ref{structure}). All these variations, in spite of a much smaller amplitude, have the same sign as the variations observed in the {\em B}- to {\em A}-DNA transition \cite{saenger}. This suggests that the twist-stretch coupling involves the same helical deformation mechanism as the transition to {\em A}-DNA. This behaviour might be important as part of the mechanism of DNA deformation which is exploited by proteins in detecting specific sequences \cite{paillardStructure2004}.

\begin{acknowledgments}
We would like to thank G. Lia and D.R. Leach for providing DNA substrates, and F. Lankas for helpful discussions. This work was supported by grants from A.R.C., C.N.R.S., the Universities Paris 6 and Paris 7, and the MOLSWITCH program.
\end{acknowledgments}

\end{document}